\documentclass[iop]{emulateapj} 
\shortauthors{Sekanina}
\shorttitle{New Kreutz Sungrazer C/2026 A1 (MAPS)} 
\slugcomment{Version \today }

\begin{document}
\title{New Kreutz Sungrazer C/2026 A1 (MAPS):\ Third Time's the Charm?}
%
%
\author{Zdenek Sekanina}
\affil{La Canada Flintridge, California 91011, U.S.A.; \it ZdenSek@gmail.com}
\begin{abstract} 
This paper describes progress achieved in early investigations of the
orbital motion and light curve of comet C/2026~A1 (MAPS), the third
ground-based discovery of a Kreutz sungrazer in the 21st century.
The highly unusual trait of the comet that has so far been ascertained
is its extraordinarily long orbital period.  The most recent orbital
computations make it increasingly likely that the object is a
fragment of one of the comets observed by Ammianus Marcellinus in
AD~363, thereby strengthening evidence in support of the contact-binary
hypothesis of the Kreutz system.  In this context, the comet is the only
second-generation fragment of Aristotle's comet that we are aware of
to appear after the 12th century.  It does not look like a major
fragment, but rather like an outlying fragment~of~a~much larger sungrazer.
In 363 it apparently separated from a parent different from the lineage of
comet Pereyra.  The light curve of comet MAPS has so far been fairly smooth,
without outbursts.~To~reach~the brightness of comet Ikeya-Seki, the comet would
have to follow an $r^{-17}$ law in the coming weeks,~which~is~unlikely.
\end{abstract}
\section{Introduction} 
\vspace{0.1cm}
Comet C/2026 A1, discovered by A.~Maury, San Pedro de Atacama,
Chile, on CCD images taken with F.~Signoret and G.~Attard as
part of the MAPS survey on 2026 January~13 (Maury 2026), is
the third Kreutz sungrazer discovered from the ground in the
21st century.  The first one, C/2011~W3 (Lovejoy), turned out to
be a ``headless wonder'' beginning a few days after perihelion,
while the second one, C/2024~S1 (ATLAS), was a dwarf sungrazer
and, in terms of expectation, a flop.  It is hoped that the
third try will be more successful.  Regardless of the
performance of the new comet near perihelion, which at
present (six weeks before perihelion) is anybody's guess,
it already achieved a new record at the time of discovery:\
it has become the Kreutz sungrazer detected at by~far~the
largest heliocentric distance preperihelion.

It is interesting to look at the history of preperihelion
discoveries of Kreutz sungrazers from the ground.  They are
a fairly exclusive club, whose definite members are the Great
September Comet of 1882 (C/1882~R1), du~Toit (C/1945~X1),
Ikeya-Seki (C/1965~S1), and the 21st century sungrazers.
According to Galle (1894), the giant 1882 comet was first
seen in the Gulf of Guinea and at the Cape of Good Hope
on 1882 September~1, 16~days before perihelion.  By sheer
coincidence, D.~du Toit discovered his 1945 sungrazer at
about the same time relative to perihelion --- 16.9~days
before it (Paraskevopoulos 1945).  K.~Ikeya and T.~Seki
caught their comet, within some 15~minutes of each other,
32.4~days before perihelion (Hirose 1965), T. Lovejoy
detected his comet 18.3~days before perihelion (Lovejoy
2011), and R.~Siverd comet ATLAS 30.9~before perihelion
(Denneau 2024).  The discovery of comet MAPS 81.4~days
before perihelion thus beats the previous record of
Ikeya and Seki by a factor of 2.5.

For the sake of completeness, I add that another sungrazer,
the Great March Comet of 1843 (C/1843~D1), is a {\it very\/}
questionable member of this group.  According to Encke
(1843), anonymous reports of the 1843 comet published in
New York newspapers mentioned that the comet was first
detected on 1843 February~5, 22~days before perihelion.
However, when introducing the new cometary designation
system in 1995, Marsden obviously did not consider this
information trustworthy enough, because if he did he
would have assigned the comet a designation C/1843~C1,
not D1.  In his comments on the classification, Marsden
(1995) did not even mention the controversy.  Additional
preperihelion observations were remarked on by Herrick
(1843) and by Peirce (1844), but dated February~19.9~UT
(Bermuda), February~24.0~UT (Philadelphia, PA), and
February~26.9~UT (Puerto Rico), they all were made less
than 10~days before perihelion.

With a touch of humor, one could also include comet Tewfik
(X/1882~K1), since its positions measured during the
total solar eclipse were according to Marsden (1967) in
good agreement with the positions that it should have
occupied 0.2~day before perihelion, if its motion followed
the orbit of the 1843 sungrazer, a plausible assumption.

The very early discovery of the new comet is of course
expected to have beneficial ramifications in terms of our
learning its properties; in this contribution I briefly
comment on two of them, which already seem to provide
exciting results.

\begin{table*}[ht] 
\vspace{0.1cm}
\hspace{-0.15cm}
\centerline{
\scalebox{1}{
\includegraphics{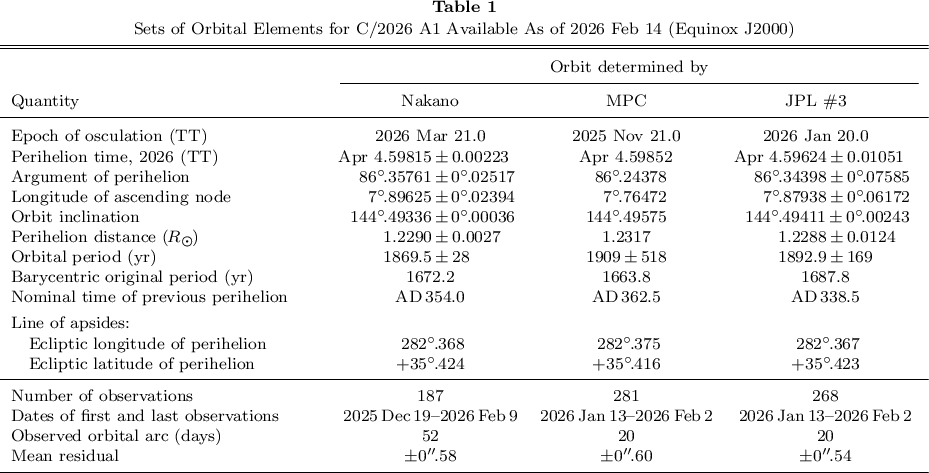}}}
\vspace{0.8cm}
\end{table*}

\section{The Orbital Elements}
\vspace{-0.05cm}
To accurately know the orbit of a comet is always important
in order to better understand the object, but~to accurately
know the orbit of a Kreutz comet is vitally important in
order to better understand the entire Kreutz sungrazer system.
The reason is the system's enormous complexity produced by its
members' susceptibility to fragmentation, especially---but not
solely---at perihelion.  Perihelion breakup, by the solar
tidal forces, dramatically affects the long-term temporal
distribution of sungrazers because every new fragment is
launched into a path with an orbital period that differs
greatly from the orbital period of a neighboring fragment.
To keep track of the total temporal distribution is not
possible, but a reliable determination of the orbital
period of any particular sungrazer allows one to constrain
its past orbital evolution and helps get insight into
broader relationships among members of the system as a
whole.

If the outcome of the MAPS comet's current, preliminary
orbital computations is confirmed in the future, when a
longer orbital arc is available, we are in for a big
surprise because the object appears to have had a very
interesting history.  I begin with comparison of the
sets of orbital elements in Table~1, available at the
time of writing (mid-February 2026).  All three have
been computed by experienced sources:\ on the left is
the updated orbit by S.\ Nakano, who has been providing
orbital data to the {\it Central Bureau of Astronomical
Telegrams\/}; his first, preliminary orbit accompanied
Green's (2026) official announcement of the comet's
discovery.  Tabulated here on the left is his improved
set (Nakano 2026),\footnote{See
{\tt http://www.oaa.gr.jp/{$\sim$}oaacs/nk.htm.}}
based on an extended orbital arc of 52~days.  In the middle
of the table is the set published in the {\it Minor Planet
Center\/}'s (MPC) Orbits/Observations Database,\footnote{See
{\tt https://minorplanetcenter.net/db\_search.}} also
available on an electronic circular (M.P.C.\,Staff 2026).  On
the right is orbit \#3, issued by the {\it Jet Propulsion
Laboratory\/}'s (JPL) Solar System{\vspace{-0.03cm}} Dynamics
Group in its {\it Small-Body Database Lookup\/} website.\footnote{See
{\tt https://ssd.jpl.nasa.gov/tools/sbdb\_lookup.html\#.}}
Unlike Nakano's orbit, the MPC and JPL orbits are based on
an orbital arc of only 20~days, as they use no prediscovery
images of the comet (for these data, see Green 2026).  This
constraint handicaps them in relation to Nakano's orbit in
that the orbital period, the element of considerable interest,
is burdened by unacceptably large uncertainty, even though
its nominal values in all three cases are fairly similar.
This is in part caused by a somewhat unfortunate circumstance
that the three sets have used different osculation epochs, but
for the orbital period this has been taken care of by computing
the barycentric values.  These are seen to agree within 24~years
or 1.5~percent, which is amazing given the short orbital arcs
and the large errors involved.

\section{The Orbital Period and Its Ramifications}
The orbital period is important because it tells us when the
object was at perihelion last time and when it will be at
perihelion next time.  Before Kreutz, the orbital periods of
sungrazing comets were unknown, which resulted in sometimes
hilarious situations with false identifications.  The periods
were dramatically underestimated by a factor of up to 20(!),
occasionally even more.  One of Kreutz's (1891, 1901) greatest
contributions was his conclusion that the {\it two\/} brightest
sungrazers of the 19th century had their periods in a general
range of 500 to 800~years.

It should be pointed out that these numbers are by no means the
lower and upper boundaries for the orbits of all Kreutz sungrazers.
For example, Marsden (1967) showed that the orbital period of
comet Pereyra when it was leaving the Solar System in 1963
was close to 900~years.  We have no idea what actually are
the orbital periods of thousands of dwarf Kreutz sungrazers
that coronagraphs on board the Solar and Heliospheric
Observatory (SOHO) and other space-borne instruments have
been detecting.  Since their majority is likely to be debris
of the Great Comet of 1106 and because their lifetimes are
known to be shorter than one revolution about the Sun, one
can infer that their orbital periods are shorter than 1000~years.

We should be thankful to comet MAPS that it is gently warning
us that we may not be necessarily correct.  All three sets of
elements in Table~1 clearly demonstrate that the comet's
orbital period is much longer than 1000~years, the number based
on Nakano's elements suggests that the period is almost certainly
longer than 1600~years.  It is the barycentric value, which
being corrected for effects of the planetary perturbations and
reduced to the barycenter of the Solar System, is the one that
counts, because it essentially measures the temporal distance
between the comet's two consecutive perihelion passages.  Thus,
the entry on line~2 minus the entry on line~8 gives the number
on line~9.  By directly integrating the comet's motion back to
the 4th century, Nakano (personal communication via D.~W.~E.~Green)
determined that perihelion was actually reached on 357 August~15.
And it is this result that makes the comet a hit.

The very long orbital period of comet MAPS is in a peculiar way
relevant to the fortunes of a contact-binary hypothesis of the
Kreutz system that I proposed a few years ago (Sekanina 2021).
If an original orbital period essentially equaling 1663~years is
confirmed, the comet will strongly support one of the fundamental
features of the hypothesis, for which evidence has been rather
tenuous.  I am referring to a group of first-generation fragments
of Aristotle's comet (assumed to be the progenitor), which were
direct parents of a number of major sungrazers in the middle
ages, including the Great Comet of 1106.  Following the
progenitor's fragmentation far from the Sun (a condition demanded
by the wide range of nodal longitudes among observed sungrazers),
the first-generation fragments were predicted to arrive at perihelion
almost simultaneously.  The hypothesis thus badly needed
evidence of not a comet, but a {\it swarm\/} of comets in Kreutz
orbits appearing at the right time --- around AD~367, midway
between 372~BC (the year of Aristotle's comet) and 1106 (the year
of the Great Comet).

The situation looked hopeless, until I picked up a book {\it The
Greatest Comets in History\/} by D.\ Seargent (2009) and glanced
through a paragraph on Roman historian Ammianus Marcellinus writing
that ``{\it in broad daylight comets were seen\/}'' in {\it late\/}
363. I had read that line elsewhere, in connection with another comet
that appeared earlier that year, moving in a path clearly inconsistent
with the expected motion of a Kreutz comet.  However, Seargent did
make the important point that the two events had nothing in common
and added several lines that absolutely astonished me.  He pointed
out that a Kreutz sungrazer could fit Ammianus' brief account and
--- as wild speculation --- the reference to {\it comets\/} (in the
plural) ``{\it may imply several sungrazing fragments close together?\/}''
In the absence of more convincing evidence this was the clincher.
What looked like {\it wild speculation\/} to the book's author,
turned out to be the {\it last piece of a puzzle\/} that I needed
to finalize my hypothesis.

The contribution from comet MAPS to this story and the provided
benefit are obvious.  While I could argue that the event, worth
no more than six words to Ammianus, was in all probability {\it
unprecedented\/}, I could not --- and had no hope to ever ---
muster up stronger arguments.  But now, all of a sudden, an object
is~\mbox{arriving} that either was seen by Ammianus from Antioch or,
more probably, is a fragment of one of the comets that~he~saw in
broad daylight.  In the least, none of the sets of elements
in Table~1 rules out this possibility, and Nakano's set~---~as
the currently most accurate one --- makes the appearance of Kreutz
comets in AD~363 increasingly likely, given that his direct determination
of the perihelion time differs by only 6$\frac{1}{4}$~years from
the critical time, with a mean error~of $\pm$28~years.  This
is my first point.

My second point is that with its very long orbital period
of 1600+~years, this object surely could hardly be one of the very
massive fragments, which must have left perihelion with only slightly
changed orbital periods, as is documented by the 1106 and 1138
sungrazers.  This means that comet MAPS provides {\it hard
evidence of tidal fragmentation\/} taking place at perihelion in
AD~363.

My third point emphasizes that thanks to its very long orbital period,
comet MAPS has been --- until this coming April --- spared a second
close approach to the Sun, so that we have the unique opportunity to
investigate a {\it second-generation fragment\/} of Aristotle's comet,
comparable, for example, to the Great Comet of 1106, albeit~on a reduced
size scale.

Two final points are related questions:\ under what conditions can
a fragment acquire such a long orbital period and what does it reveal
about comet MAPS' direct parent (that is, a first-generation fragment)
and its identity.  I address these problems in detail below.

\section{Orbital Period As a Product of\\Conditions At Separation}
The fragmented nucleus of the Great September Comet of 1882 resembled
after perihelion a line of {\it small beads strung on a thread of worsted\/},
as Gill (1883) put~it.  This appearance, easily resolved a few weeks
after separa\-tion, meant that the individual masses must have ended
up in {\it very\/} different orbits.  Indeed, in Part~II of his famous
treatise {\it Untersuchungen \"{u}ber das System der Cometen 1843~I,
1880~I und 1882~II\/}, Kreutz (1891) computed that the orbital periods
of four brightest nuclear fragments ranged over an interval of almost
300~years, from about 670~years to nearly 960~years.

The reason for the large variations in the orbital period is simple:
the orbital velocities of sungrazers are just below the escape velocity
from the Sun, when the orbital period becomes infinitely long.  For
example, the 1882 sungrazer in an orbit with a period of 772~years had
at perihelion a velocity approximately 11~m/s below the escape limit.
To end up in an orbit with a period of 872~years, a fragment only needed
to acquire a separation velocity of 0.86~m/s (for example, rotational
in nature) in the direction of the orbital velocity.

There is another way to get a fragment into an orbit with a
period of 872~years, namely by placing~it~at~peri\-helion 4.16~km
farther from the Sun along the radius vector than was the 1882 sungrazer
(orbiting with the same period of 772~years as before) and keeping the
orbital velocity unchanged.

In reality this is what happens when a fragment breaks off from its
parent at rest.  The distance of 4.16~km is in fact the separation
distance of the center of mass of the fragment from the center of
mass of the parent along the radius vector, as illustrated in Figure~1.

\begin{figure}[ht] 
\vspace{0.15cm}
\hspace{-0.2cm}
\centerline{
\scalebox{0.635}{
\includegraphics{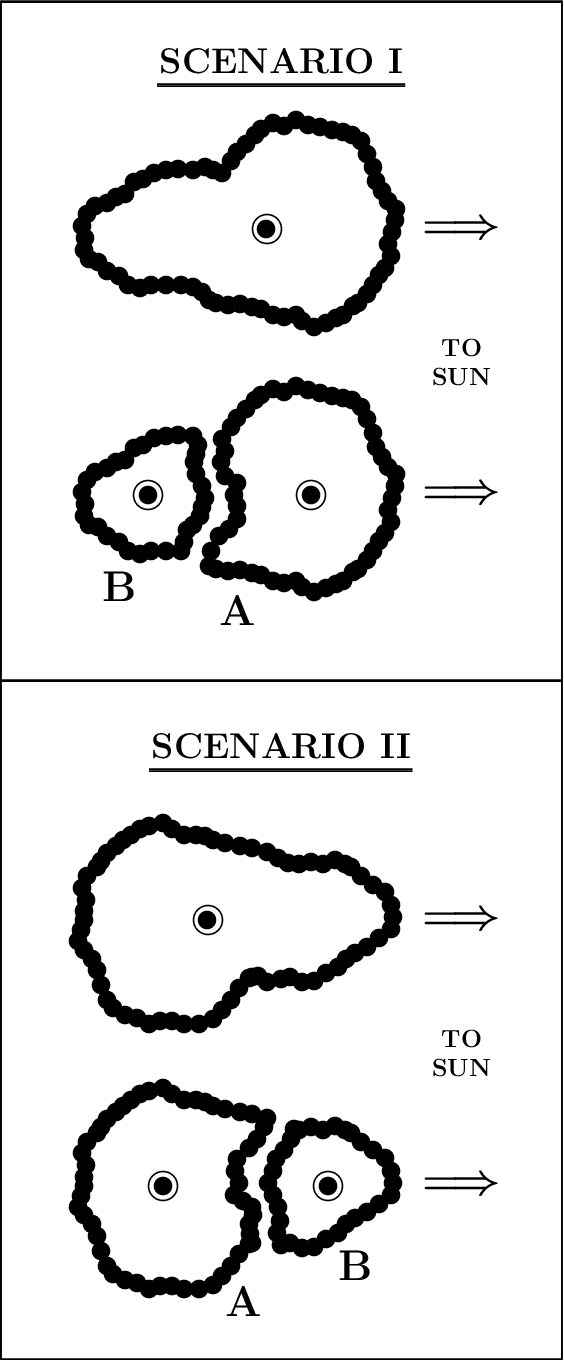}}}
\vspace{0cm}
\caption{A sungrazer's nucleus in close proximity of perihelion
shortly before and after breaking up tidally into two uneven
fragments.  In Scenario~I (top) it is the more sizable fragment~A
that begins its existence on the sunward side of the parent's
nucleus (to the right).  It ends up in an orbit of shorter period
than the parent's.  On the other hand, the smaller fragment B, on
the far side from the Sun, enters an orbit of longer period than
was the parent's --- the case of comet MAPS.  In Scenario~II, the
positions of the fragments are reversed and so are their
future orbital periods. The circled disks are the centers of mass
of the parent comet and the two fragments. (Adapted from Sekanina
\& Kracht 2022.){\vspace{0.2cm}}}
\end{figure}

Under the effects of the solar tidal forces there are thus two ways
to get fragments into orbits with widely different periods:\ by momentum
exchange (via a separation velocity) or by virtue of shifts in the
center of mass along the radius vector with no momentum change.
Pursuing this latter option:\ if $P_{\rm par}$ is the orbital period of
the parent, $P_{\rm frg}$ the orbital period of the fragment, and $r_{\rm
frg}$ the heliocentric distance at the time of separation, the distance
$u_{\rm frg}$ of the fragment's center of mass from the parent's center
of mass along the radius vector can be expressed as
\begin{equation}
u_{\rm frg} = {\textstyle \frac{1}{2}} r_{\rm frg}^2 \!\left( \!
 P_{\rm par}^{-\frac{2}{3}} \!-\! P_{\rm frg}^{-\frac{2}{3}} \!\:\!
 \right)\!,\\[-0.1cm]
\end{equation}
where $P_{\rm par}$ and $P_{\rm frg}$ are in years and $r_{\rm frg}$
and $u_{\rm frg}$ in~AU.  The minimum distance $u_{\rm frg}$ is
obtained, when~the~frag\-mentation event is assumed to take place
exactly at peri\-helion, \mbox{$r_{\rm frg} = q_{\rm par}$}.

Coming back to comet MAPS, the minimum distance $u_{\rm frg}$ can readily
be determined, if we know its parent's orbital period and perihelion
distance.  Since the birth of the Kreutz system at large heliocentric
distance had essentially no effect on the orbital periods of the
first-generation fragments and they all arrived at perihelion over
a very short period of time, \mbox{$P_{\rm par} = 735$ years}.  With
\mbox{$P_{\rm frg} = 1663$ years}, one finds from Equation~(1)
\begin{equation}
u_{\rm frg}({\rm MAPS}) = 8.336 \, q_{\rm par}^2,
\end{equation}
where $q_{\rm par}$ is now in solar radii and $u_{\rm frg}$ comes out
in km.  Since the ratio of the perihelion distances of the fragment
to the parent, \mbox{$(q_{\rm par} \!+\! u_{\rm frg})/q_{\rm par}
\rightarrow 1$}, all one needs to do is to integrate the motion of
comet MAPS back to AD~363 in order to obtain $q_{\rm par}$.  Nakano
(personal communication via D.~W.~E.~Green) thus obtained \mbox{$q_{\rm par}
= 1.187\:R_\odot$} and
\begin{equation}
u_{\rm frg}({\rm MAPS}) \simeq 11.7\;{\rm km}.
\end{equation}

The conclusion is that if the radial distance of the center of mass
of the fragment to become comet MAPS from the center of mass of its
parent was about 12~km, the fragment would have ended up in an orbit
with a period of 1663~years, if it separated exactly at perihelion.  Thus,
the unusually long orbital period of comet MAPS is readily understood
as that of an outlying fragment (at far left in Scenario~I of Figure~1)
of a sungrazer{\vspace{-0.05cm}} more~than $\sim$20~km across.  But since
\mbox{$u_{\rm frg} \sim r_{\rm frg}^2$}, it{\vspace{-0.063cm}} is highly
sensitive to the point of separation in the orbit, the actual distance
also depends on the angle with the radius~vector.

\section{Relationship to Major Kreutz Sungrazers:\\Preliminary Results}
In a classification system that I introduced~for~Kreutz sungrazers
(e.g., Sekanina 2022), comet MAPS appears to belong to Population~Pe,
because its nodal longitude is currently nearly identical with that
of comet Pereyra (C/1963~R1).  However, an in-depth investigation is
in order to rule out coincidence.

The straightforward link of comet MAPS to the apparition of AD~363
provides us with the unique opportunity to examine the level of its
relationship with other Kreutz sungrazers, the major ones in particular,
at that time.  Table~2 compares critical orbital parameters of comet
MAPS (or its parent) with the lineage of 363--1106--1843 (Population~I),
363--1138--1882 (Population~II), as well~as 363--1041-1963 (Population~Pe).
It is apparent that comet MAPS had its own parent, although it derived
from Lobe~I, the source of Populations~I and Pe.  The numbers for Pe
depend on whether the link of Pereyra to the September 1041 comet is
correct.  Given that the orbit of comets MAPS is still in the process
of being refined, the present conclusions are preliminary.

\begin{table}[b] 
\vspace{0.6cm}
\hspace{-0.2cm}
\centerline{
\scalebox{0.98}{
\includegraphics{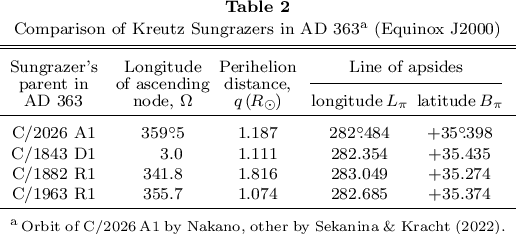}}}
\vspace{-0.07cm}
\end{table}

\section{The Light Curve}

Light curve is another cometary characteristic that greatly benefits
from early discovery.  I already showed in Section~1 that there is a
total of only six Kreutz sungrazers discovered from the ground before
perihelion.  Their number with a meaningful preperihelion light curve is
even smaller.  In the following I present two plots, the first of which
(Figure~2) displays a normalized total magnitude $H_{\:\!\!\Delta\:\!\!}
(r, 1, 0^\circ\!, {\sf corr})$ as a function of time from perihelion
(negative numbers mean before perihelion), the other (Figure~3)
exhibits the normalized magnitude as a function of heliocentric
distance $r$ (in AU).
\begin{figure*}[ht] 
\vspace{0.05cm}
\hspace{-0.18cm}
\centerline{
\scalebox{0.94}{
\includegraphics{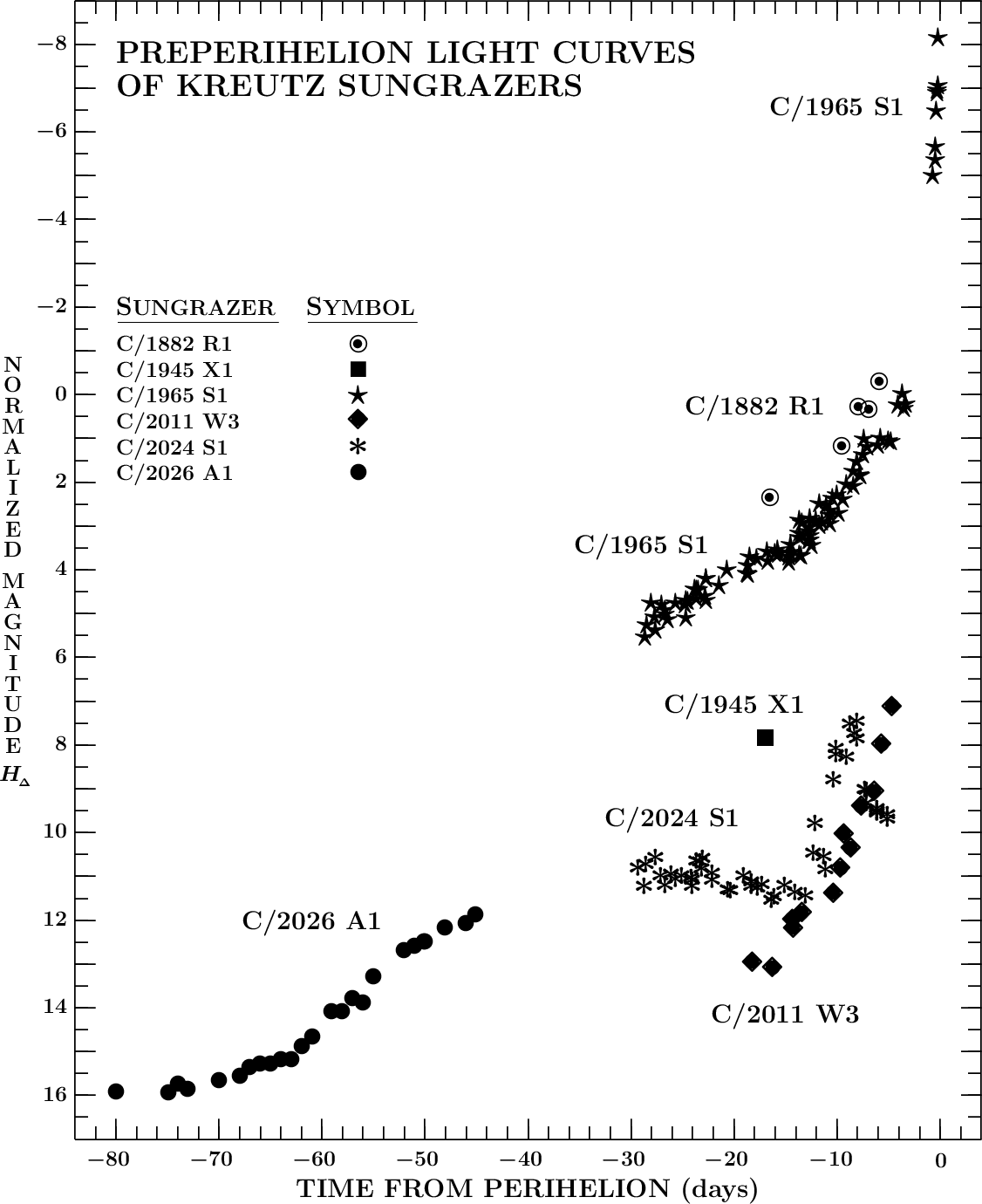}}}
\vspace{-0.1cm}
\caption{Preperihelion light curves of the Kreutz sungrazers
discovered from the ground.  Observed magnitudes were taken
mostly from the {\it Comet Observation Database\/} (COBS), for
C/1882~R1, C/1945~X1, and C/1965~S1 from other sources, as
mentioned in the text.  The normalized magnitude has been
reduced to 1~AU from the Earth by an inverse square power
law, to a zero phase angle by the Marcus law, and standardized
by applying the personal/instrumental correction.  Among the
interesting features of the plot is that C/2024~S1, which
did not survive perihelion, was mostly brighter than C/2011~W3,
which did survive it (though barely), and that C/2026~A1 will
be substantially fainter near perihelion than C/1965~S1 unless
its rate of brightening picks up a lot between 40 and 30~days
before perihelion.}
\end{figure*}
\begin{figure*} 
\vspace{0.05cm}
\hspace{-0.18cm}
\centerline{
\scalebox{0.94}{
\includegraphics{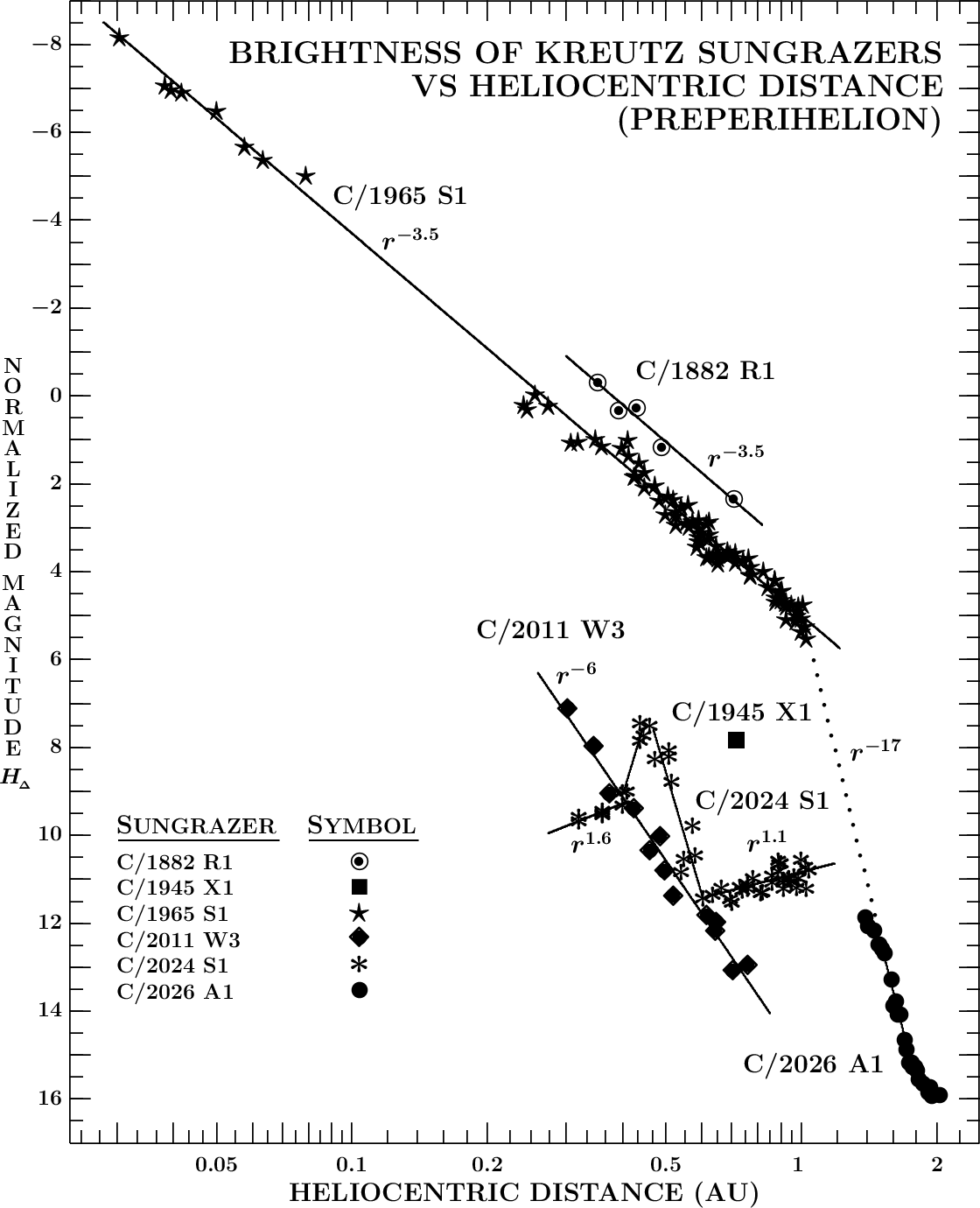}}}
\vspace{-0.1cm}
\caption{Dependence of the normalized magnitude of the Kreutz
sungrazers from Figure 2 on heliocentric distance before perihelion.
For the sources of the data, see the caption to Figure~2 and text.
Interesting features of the plot include the same $r^{-3.5}$
law that C/1882~R1 and C/1965~S1 follow, even though the data for
the first object are scanty and the slope uncertain; C/2011~W3
brightened much more rapidly; C/2024~S1 was actually fading on
its approach to the Sun both before and{\vspace{-0.045cm}} after
the outburst; and for C/2026~A1 to reach the brightness of
C/1965~S1, it would have to brighten according to an $r^{-17}$
law at heliocentric distances between 1.4~AU and 1~AU.}
\end{figure*}
The normalized magnitude $H_{\:\!\!\Delta\:\!\!}(r,1,0^\circ\!,{\sf
corr})$ was calculated from the reported total apparent magnitude
$H(r,\Delta,\alpha)$, where $\Delta$ is the geocentric distance
(in AU), $\alpha$ is the phase angle, and {\sf corr} is a constant
correction that depends on the observer, instrument, and magnitude
type (visual, CCD, filter, etc.).  The relation between the
normalized and apparent magnitudes is
\begin{equation}
H_{\!\Delta\!}(r,1,0^\circ\!,{\sf corr}) \!=\!
 H(r,\Delta,\alpha) - 5 \log \Delta - 2.5 \log \Phi(\alpha) + {\sf corr},
\end{equation}
where $\Phi(\alpha)$ is the phase law developed by Marcus (2007) for
dust-poor comets.

I now discuss the light-curve data for each of the six objects
in Figures~2 and 3 in chronological order.  The magnitudes of
the Great September Comet of 1882 were very difficult to locate.
The first one, for the discovery date of 1882 September~1 is
not the comet's brightness estimate at all; because the sightings
were with the naked eye by non-astronomers, it was assumed that
the apparent brightness was the same as for any other unknown
comet detected accidentally for the first time.  I used a
magnitude of 3.5, which Broughton (1979) obtained for Halley's
comet from the historical appearances before its periodicity
was discovered.  The adopted value probably still is an
underestimate of the actual brightness because the 1882 comet
was then only 30$^\circ$ from the Sun.

The second entry for the 1882 sungrazer was provided by W.\ H.\ Finlay,
an observer at the Royal Observatory, Cape of Good Hope, and reported
by Gill (1882).  The account, referring to September~8, did not offer
a total magnitude because it specifically mentioned that the {\it
nucleus\/} appeared to the naked eye as bright as a star of the
3rd~magnitude.  The observation was assigned a correction of $-$1~mag
in an effort to allow for the brightness difference between the
nuclear condensation and the whole coma.

The last three magnitude estimates of the 1882 sungrazer originated
from New Zealand and were reported by Orchiston et al.\ (2020):\
from Wellington on September~9, when the nucleus was compared to
a star of the 2nd~magnitude; from a coastal collier crossing
the Bay of Plenty on September~10, when the nucleus was again
equated to a star of the 2nd~magnitude in brilliancy; and from
Gisborne on September~11, when the brilliant nucleus was almost
equal to a star of the 1st~magnitude.  Thus, in each case it
was clearly stated that the brightness estimate referred to the
nucleus, which was always described as very, unusually, or
particularly brilliant.  As with Finlay's magnitude, I rather
arbitrarily assigned a correction of $-$1~mag to each of the
New Zealand estimates.

The next sungrazer in Figures~2 and 3, discovered by du Toit
(C/1945~X1), was taken care of easily, because there was only
a single brightness estimate from the discovery plate
(Paraskevopoulos 1945).  The data point was plotted with no
correction applied, although there obviously is an color-index
effect.

The preperihelion light curve of the famous comet Ikeya-Seki
(C/1965~S1) has been covered very well, with only a 2.7~day gap
around 2~days before perihelion.  The curve plotted has been
copied from my earlier paper (Sekanina 2002), except that the
phase effect has now been incorporated.  Because backscattering
dominated until about 7~days before perihelion (0.4~AU) and
forward scattering nearer perihelion, the overall effect was to
diminish the rate of brightening (to $r^{-3.5}$) and to make
the preperihelion absolute magnitude brighter (close to
magnitude~5).  I am not going to detail the brightness data
here, but only remark that there were obtained by 23~observers
using 29~instruments; 10~observers provided naked-eye estimates.
The total number of used magnitude observations was 92.

The light curve for comet Lovejoy consists of observations by
three observers:\ J.\ \v{C}ern\'y (31-cm Schmidt-Cassegrain,
CCD magnitudes, V~filter), T.\ Lovejoy (20-cm Schmidt-Cassegrain,
CCD magnitudes on CBET 2930), and G.~W.~Wolf (18-cm reflector,
visual magnitudes).  The observers' magnitude scales were found
to be mutually consistent and a correction of $-$0.4~mag was
applied throughout.  The light curve is similar to that
presented in Sekanina \& Chodas (2012) and clearly much steeper
than the light curves of the intrinsically bright sungrazers.
A total of 13 data points.

The light curve for comet ATLAS is based on data by three
observers as well:\ M.\ Ma\v{s}ek (6.7-cm lens, CCD magnitudes,
V~filter, correction $-$0.3~mag; 10.7-cm lens, CCD magnitudes,
V~filter, correction $-$0.45~mag; 30-cm Cassegrain, CCD magnitudes,
V~filter, correction $-$0.7~mag), M.~Mattiazzo (7-cm lens,
CCD magnitudes, correction $-$0.7~mag; 25-cm reflector, CCD
magnitudes, correction $-$1.3~mag); and A.~R.~Pearce (5-cm
refractor, CCD magnitudes, no or V~filter, correction $-$1.2~mag;
10.6-cm refractor, CCD magnitudes, correction $-$1~mag).  The
light curve is obviously unlike any other in the plot and could
serve for the future as a standard for early identification of
a dwarf sungrazer:\ outbursts and the brightness stalling if
not decreasing (\mbox{$n \leq 0$} in the $r^{-n}$ law) in
between.  A total of 50 observations.

Finally, a preliminary light curve for comet MAPS is in the
plots represented by Pearce's 25 CCD magnitudes made with the
35-cm and 43-cm Schmidt-Cassegrain reflectors, the applied
correction being $-$1~mag.  The purpose for presenting
Figures~2 and 3, the light curve of comet MAPS reveals the
behavior of a Kreutz sungrazer long before perihelion.  Both
plots show an accelerated brightening, which seems to have
ceased more than 50~days before perihelion.  It remains to
be seen whether the slowdown in Figure~2 is permanent or
temporary.  An interesting feature in Figure~3 is the
illustration that to become as bright as Ikeya-Seki, the
comet would have to keep up with an $r^{-17}$ rate of brightening
between 1.4~AU and 1~AU, which at present does not look
sustainable.

\section{Final Remarks and Conclusions}
This paper is essentially a progress report on the nature,
history, and behavior of comet MAPS and its possible place
in the hierarchy of the Kreutz sungrazer system, based on
a preliminary investigation of its orbit and light curve.
Some of the conclusions may have to be modified in the
future, as additional data become available.

At present, the most unexpected feature is the comet's extraordinarily
long orbital period, at least twice as much as the orbital
periods of other Kreutz sungrazers with well-determined
orbits.  The high quality of astrometric observations has
made it possible to establish the orbital period with
a standard deviation better than 2~percent, from an arc
of 52~days long, thanks to numerous, promptly evaluated
prediscovery observations.

Although this result still needs to be further refined by
updated computations based on observations covering longer
orbital arcs, it has to be remembered that as the comet
gets closer to the Sun, its motion is likely to get affected
by nongravitational forces perceptible enough to introduce
uncertainties caused by their imperfect modeling.  The degree
of accuracy with which the orbital period is eventually
determined remains to be seen.

At present, Nakano's computations give for the nominal time
of the previous perihelion mid-August 357, with a standard
error of merely $\pm$28~years.  The nominal time differs by
only 6$\frac{1}{4}$~years from Ammianus Marcellinus' 
reported sighting of {\it comets in broad daylight\/} in
late 363.  In the context of the contact-binary hypothesis
that I introduced in September 2021, a swarm of sizable
fragments was expected to arrive at perihelion midway
between Aristotle's comet of 372~BC and the Great Comet
of 1106 to fit the pattern completed by the 1843 sungrazer,
and the brief account by Ammianus fits that expectation
perfectly.  If the definitive original orbital period
turns out to essentially equal 1663~years, we will have hard
evidence in support of the scenario.

Other important points in the same context suggest that
comet MAPS (i)~is the only second-generation fragment of
Aristotle's comet that we are aware of to appear after the
12th century; and, given its long orbital period, which
essentially rules out the status of a major fragment,
and the predicted perihelion distance of its parent in 363,
(ii)~is probably an outlying fragment of a sungrazer
more than $\sim$20~km in diameter.

In the hierarchy of the Kreutz sungrazer system, comet
MAPS does not appear to be closely related to comet
Pereyra, as it may seem because of the coincidence of
their nodal lines.  Nakano's orbital computations for
for comet MAPS suggest that its parent differed from
Pereyra's if the latter's ancestry involved
the September comet of 1041 (Sekanina \& Kracht 2022).

Preperihelion light curves of several Kreutz sungrazers
from ground-based observations display a wide variety of
behavior.  The light curve of comet MAPS, which is{\pagebreak}
the first Kreutz sungrazer observed more than 33~days
before peri\-helion, suggests an accelerated brightening
with time at larger heliocentric distances, which
appears to cease more than 50~days before perihelion.
It remains to be seen whether this change is permanent
or temporary.  There have been no outbursts.  In order
for the comet to be eventually as bright as Ikeya-Seki,
the rate of brightening between 1.4~AU and 1~AU from
the Sun would have to follow an $r^{-17}$ law, an
unlikely trend.\\[-0.1cm]

I thank S.~Nakano for providing me with the results of
his unpublished computations and D.~W.~E.~Green for
his help in facilitating communication.\\[-0.4cm]

\begin{center}
{\footnotesize REFERENCES}
\end{center}
\vspace{-0.45cm}
\hspace{-1cm}
\begin{itemize}
{\footnotesize
\item[\hspace{-0.3cm}]
\hspace{-0.88cm}Broughton, R.\ P.\ 1979, J.\ Roy.\ Astron.\ Soc.\ Canada, 73, 24
\\[-0.57cm]
\item[\hspace{-0.3cm}]
\hspace{-0.88cm}Denneau, L.\ 2024, CBET 5453 
\\[-0.57cm]
\item[\hspace{-0.3cm}]
\hspace{-0.88cm}Encke, J.\ F.\ 1843, Astron.\ Nachr., 20, 349
\\[-0.57cm]
\item[\hspace{-0.3cm}]
\hspace{-0.88cm}Galle, J.\ G.\ 1894, Cometenbahnen. Leipzig:\ Verlag W.\ Engelmann,{\linebreak}
 {\hspace*{-0.6cm}}315pp
\\[-0.57cm]
\item[\hspace{-0.3cm}]
\hspace{-0.88cm}Gill, D.\ 1882, Mon.\ Not.\ Roy.\ Astron.\ Soc., 43, 19
\\[-0.57cm]
\item[\hspace{-0.3cm}]
\hspace{-0.88cm}Gill, D.\ 1883, Mon.\ Not.\ Roy.\ Astron.\ Soc., 43, 319
\\[-0.57cm]
\item[\hspace{-0.3cm}]
\hspace{-0.88cm}Green, D.\ W.\ E.\ 2026, CBET 5658
\\[-0.57cm]
\item[\hspace{-0.3cm}]
\hspace{-0.88cm}Herrick, E.\ C.\ 1843, Amer.\ J.\ Sci.\ Arts, 45, 188
\\[-0.57cm]
\item[\hspace{-0.3cm}]
\hspace{-0.88cm}Hirose, H.\ 1965, IAUC 1921
\\[-0.57cm]
\item[\hspace{-0.3cm}]
\hspace{-0.88cm}Kreutz, H.\ 1891, Publ.\ Sternw.\ Kiel, 6
\\[-0.57cm]
\item[\hspace{-0.3cm}]
\hspace{-0.88cm}Kreutz, H.\ 1901, Astron.\ Abhandl., 1, 1
\\[-0.57cm]
\item[\hspace{-0.3cm}]
\hspace{-0.88cm}Lovejoy, T.\ 2011, CBET 2930
\\[-0.57cm]
%
%
\item[\hspace{-0.3cm}]
\hspace{-0.88cm}Marcus, J.\ N.\ 2007, Int.\ Comet Quart., 29, 39
\\[-0.57cm]
\item[\hspace{-0.3cm}]
\hspace{-0.88cm}Marsden, B.\ G.\ 1967, Astron.\ J., 72, 1170
\\[-0.57cm]
\item[\hspace{-0.3cm}]
\hspace{-0.88cm}Marsden, B.\ G.\ 1995, Int.\ Comet Quart., 17, 3
\\[-0.57cm]
%
%
%
%
\item[\hspace{-0.3cm}]
\hspace{-0.88cm}Maury, A.\ 2026, CBET 5658 
\\[-0.57cm]
\item[\hspace{-0.3cm}]
\hspace{-0.88cm}Minor Planet Center Staff 2026, MPEC 2026-C98
\\[-0.57cm]
\item[\hspace{-0.3cm}]
\hspace{-0.88cm}Nakano, S.\ 2026, NK 5553
\\[-0.57cm]
\item[\hspace{-0.3cm}]
\hspace{-0.88cm}Orchiston, W., Drummond, J., \& Kronk, G.\ 2020, J.\ Astron.\ Hist.{\linebreak}
 {\hspace*{-0.6cm}}Herit., 23, 628
\\[-0.57cm]
\item[\hspace{-0.3cm}]
\hspace{-0.88cm}Paraskevopoulos, J.\ S.\ 1945, IAUC 1024
\\[-0.57cm]
\item[\hspace{-0.3cm}]
\hspace{-0.88cm}Peirce, B.\ 1844, American Almanac.  Boston:\ J.\ Munroe, 94
\\[-0.57cm]
\item[\hspace{-0.3cm}]
\hspace{-0.88cm}Seargent, D.\ 2009, The Greatest Comets in History:\ Broom Stars{\linebreak}
 {\hspace*{-0.6cm}}and Celestial Scimitars. New York:\ Springer
 Science+Business{\linebreak}
 {\hspace*{-0.6cm}}Media, 260pp
\\[-0.57cm]
%
%
\item[\hspace{-0.3cm}]
\hspace{-0.88cm}Sekanina, Z.\ 2002, Astrophys.\ J., 566, 577
\\[-0.57cm]
\item[\hspace{-0.3cm}]
\hspace{-0.88cm}Sekanina, Z.\ 2021, eprint arXiv:2109.01297
\\[-0.57cm]
\item[\hspace{-0.3cm}]
\hspace{-0.88cm}Sekanina, Z.\ 2022, eprint arXiv:2212.11919
\\[-0.57cm]
\item[\hspace{-0.3cm}]
\hspace{-0.88cm}Sekanina, Z., \& Chodas, P.\ W.\ 2012, Astrophys.\ J., 757, 127
\\[-0.65cm]
\item[\hspace{-0.3cm}]
\hspace{-0.88cm}Sekanina, Z., \& Kracht, R.\ 2022, eprint arXiv:2206.10827}
%
%
\vspace{-0.23cm}
\end{itemize}
\end{document}